\documentclass[11pt]{article}

\usepackage[top=50mm, bottom=50mm, left=50mm, right=50mm]{geometry}
\usepackage{lineno}
\usepackage{amssymb}
\usepackage{amsmath}
\usepackage{amsthm}
\usepackage{epsfig}
\usepackage{graphicx}
\usepackage{graphics}
\usepackage{float}
\usepackage{subfigure}
\usepackage{multirow}
\usepackage{color}
\usepackage{lineno}
\usepackage{fullpage}
\usepackage[normalem]{ulem} 
\usepackage{makeidx}
\usepackage{xspace}
\usepackage{wrapfig}
\makeindex

\definecolor{darkred}{rgb}{1, 0.1, 0.3}
\definecolor{darkblue}{rgb}{0.1, 0.1, 1}
\definecolor{darkgreen}{rgb}{0,0.6,0.5}

\newcommand {\mm}[1] {\ifmmode{#1}\else{\mbox{\(#1\)}}\fi}

\begin{document}

\title{To Mock a Mockingbird: Studies in Bio-mimicry}
 
\author{
Inavamsi Enaganti\footnote{Courant Institute of Mathematical Sciences, New York University, Email : ibe214@nyu.edu} \and Bud Mishra\footnote{Courant Institute of Mathematical Sciences, New York University, Email : mishra@nyu.edu}
}

%\authorrunning{Mokhov, Sutcliffe and Voronkov}

% \title{Some Interesting results in Biomimicry}
% \author{Inavamsi Enaganti} \and \author{Bud Mishra}
%\date{}

%\begin{document}
\maketitle
%\linenumbers
\setcounter{page}{0}

\begin{abstract}
This paper dwells on certain novel game-theoretic investigations in bio-mimicry, discussed from the perspectives of information asymmetry, individual utility and its optimization via strategic interactions involving co-evolving preys (e.g., insects) and predators (e.g., reptiles) who learn. Formally, we consider a panmictic ecosystem, occupied by species of prey with relatively short lifespan, which evolve mimicry signals over generations as they interact with predators with relatively longer lifespans, thus endowing predators with the ability to learn prey signals.  Every prey sends a signal and provides utility to the predator. The prey can be either nutritious or toxic to the predator, but the prey may signal (possibly) deceptively without revealing its true ``type.'' The model is used to study the situation where multi-armed bandit predators with zero prior information are introduced into the ecosystem. As a result of exploration and exploitation the predators naturally select the prey that result in the evolution of those signals. This co-evolution of strategies produces a  variety of interesting phenomena which are subjects of this paper.

\end{abstract}

\textbf{Keywords :}
Evolutionary Game, Signaling Game, UCB, multi-armed bandit, Predator-Prey, Mimicry, prey toxicity, learning predator, evolving prey, counter-intuitive results

\section{Introduction}

Standard predator-prey models treat predators and prey as non-evolving, non-learning entities. In reality, over time both predators and prey start a co-evolutionary chase which is fueled by continual evolution and learning. A simple example is that of a literal chase where the predators consume the slower prey which are the easiest to catch. The remaining prey population is faster on average and succeeding generations of prey are naturally selected to be faster in a Darwinian evolution. As a result of quicker prey, under negative selection pressure, the slower predators are pruned out over time, which results in faster predators. It's the beginning of an adversarial chase!

Similarly toxicity of prey and resistance of predator evolve through generations, if we consider the case where the lifespan of  a predator covers a significantly large number of generations of prey. Such a co-evolution may be further limited by the premise that it will not let the predator gain ability to evolve fast enough to develop immunity to the toxic prey. But instead, as seen in nature, we may endow the predators with a learning mechanism and thus allow them to learn and continually adapt over time.[1] When a predator consumes a prey they update their inner beliefs about the prey based on the effect the prey has on the predator. To this extent, this paper treats the predator as a learning agent who uses reinforcement learning to update their policy of choosing to eat or not eat a prey of the same type when encountering it again in the future. But to effectively execute such strategies, a necessary requirement is the ability of the predator to distinguish between multiple prey types so that they can update the respective belief regarding that type. This ability to distinguish arises from the signal sent by the corresponding prey. A signal can be visual, acoustic or even a smell. For example a red bird sends a visual signal, different from a  blue bird.[2] \\

Breaking it down, the predators have prior beliefs which may be learnt during the predator's lifetime, or genetically programmed. Based on the signal the predator receives from the prey they choose to either consume or avoid the prey that sent the signal. Due to prey evolution and natural selection the signals of a prey species changes through the generations. It gets even more interesting when two different species evolve to send the same signal. Effectively the predator cannot distinguish between the two and will thus share a common belief with respect to the common signal. This is commonly known as prey mimicry.[3] Some common types of mimicry are Batesian and Mullerian mimicry which arise due to prey toxicity. For a non evolving predator-prey system one can find equilibrium conditions and dynamics but it gets interesting in our current situation of learning predators and evolving prey.[4] \\

The model has many direct applications to vaccine design that mimics a yet-to-be-encountered (but known) virus. Guided by the model, the designed vaccine embeds itself in a Mullerian mimicry-ring, thus developing tolerance from and memory in the host-immune system. A quasi-Batesian virus then has difficulties to invade the host despite its previously-acquired abilities to deceive the host system via mimicry. As the adversarial chase ensues, virus evolves additional variants, which need to be recognized, analyzed and learned (AI/ML), leading to booster vaccines, resulting in a complex game of Whack-a-Mole. What would be the best learning strategies, if one's goal is to achieve a herd immunity rapidly in the population with respect to the collection of variants?   \\

Thus, we see a fascinating variety of phenomena arising due to differences in efficacy-values the predator (resp. vaccine) has with respect to the different prey-species (resp. virus). We effectively model this using utility. For example, the utility of a toxic prey to a predator will be negative while a nutritious prey will be positive. Furthermore the utility function encodes preference and is exactly what we need to model the behaviour of evolving predators and prey. In regards to two mimicking species that share a common signal, the utility the predator associates to the signal will be the expected utility from the two species in question.\\

To keep matters simple, we analyse the following somewhat idealized system. We consider a signalling game with multiple species of predator and prey. The prey (Lifespan of Mayfly $\approx$24hours) has a relatively short lifespan compared to the the predator (Lifespan of Common Toad $\approx$12years). This way the prey can undergo mutations in its signal and adapt while the predator can only learn about the prey but cannot change genetically by germ-line evolution in the short period. As the signals of the prey mutate and evolve there will be convergence and divergence of signals of different species. We try to understand the conditions and extent of mimicry with respect to the utility value of different prey species to the predator. Here, the predators will all be modeled as multi-armed bandit agents. This dynamics results in a few interesting phenomena that arise as a mathematical consequence of our assumptions.\\

%\paragraph{New work.}
\vspace*{0.08in}\noindent{\bf New work}\\
Bio-mimicry has scarcely been interpreted using a formal mathematical framework of utility. Another novel idea is using multi-armed bandit predators that can continuously learn and update their belief regarding the non-stationary distribution of multi-dimensional signals. \\

Most of the following phenomena explained below are just mathematically feasible scenarios. Certain phenomenon such as: `Mimicry can be bad to both species' is discussed here for the first time, albeit, such hypothetical scenarios may or may not have natural/biological feasibility. Thus, `Mimicry can be bad to both species' may only refer to a novel but refutable hypothesis suggesting that mimicry is bad resulting in the extinction of both species involved, and will require falsifying experiments.\\

\section{Model}

\subsection{Ecosystem}

The ecosystem consists of multiple prey and multiple predators. On a given day a predator encounters some number of prey. Each prey gives out a respective signal. Based on the predators prior knowledge it chooses to consume a prey emitting a specific signal. After eating the prey, based on the utility it got from the prey it updates its information about the signal. This continues over time leading to a variety of phenomena. \\

It is to be noted that the prey mutate and evolve over each generation but the predator whose lifespan is significantly longer than the prey continuously learns and updates its knowledge. As naturally seen, the predator comes across a randomly sampled set of prey and chooses to consume one prey out of them. \\

There are two types of ecosystems we can focus on. In the first the prey also compete with each other for resources. In the second one the prey have a reproduction rate and can indefinitely grow if there is no predation. Both ecosystems have very distinct properties and showcase very different phenomena under similar conditions. We mainly focus on the indefinite growth in the examples seen in this paper.

\subsection{Prey}
Every prey has the following properties which will be defined below.
\begin{itemize}
    \item Gene Vector, which encodes toxicity, evasion rate, etc.
    \item Signal, which is function of the gene vector
    \item Mutation rate
    \item Reproduction rate
\end{itemize}

\paragraph{Species}
Prey divide into species. Every prey of a specific species share common traits such as reproduction rate, mutation rate and genetic constraints. In the paper each species will be represented by a different color (signaling scheme).

\paragraph{Gene vector, Signal and Utility to Predator}
Every prey has some genetic values, which we can encode as a gene vector. The gene vector is bound by the genetic constraints of the prey species. A natural analogy is the set of chromosomes and the genomic information therein. The allelic values assumed by the  set of genes in the gene vector determine a variety of traits -- from how the prey looks to its speed of evasion from the predator. It essentially is analogous to a change in chemical composition which changes the taste, toxicity and nutritional value to the predator. To sum it up, the gene vector determines the signal the prey sends and also the utility a predator receives from consuming the respective prey.\\

We consider a signaling range bounded in the range $[L,R]$. Let $g$ be the signal function that takes gene vector as input and returns signal.

Example 1: Consider two prey species $A$ and $B$. Let the gene vector of $A$ be [2,1] and the gene vector of $B$ be [1,2]. Suppose $g([x,y])=x+y$ is the signaling mechanism of both species, then  prey both $A$ and $B$ give out the same signal 3. Given a predator $P$ with utility function $f([x,y])=x^2-y$, it can be seen that the utility of $A$ to the predator is 3 while the utility of $B$ to the predator is -2 which implies that $B$ is toxic to the predator $P$. $P$ cannot distinguish between the perfect mimics $A$ and $B$ as it receives the same signal 3. \\

Example 2: Consider two prey species $A$ and $B$. Let the gene vector of $A$ be [2,1] and the gene vector of $B$ be [1,2]. Suppose $g([x,y])=2x+y$ is the signaling mechanism of the species $A$ and $g([x,y])=x+y$ is the signaling mechanism of the species $B$. Then prey of $A$ signal 5 while $B$ signals 3. Given a predator $P$ with utility function $f([x,y])=x^2-y$, it can be seen that the utility of $A$ to the predator is 3 while the utility of $B$ to the predator is $-2$ which implies that $B$ is toxic to the predator $P$. $P$ can clearly distinguish between the two prey and will learn that signal 5 corresponds to a good prey while signal 3 corresponds to bad prey. \\

Since the signal is the only trait visible to the predator, we look specifically at parts of the gene vector affected by the signal. For the sake of simplicity we assume that the signaling function ($g$) to be a injective function. This assumption implies that the gene vector can be inferred from the signal. As the signaling function is injective, we can assume that the utility to the predator which depends on gene vector depends on the signal. We thus use the notation $u_{AP}(s)$ which refers to the utility of prey $A$ with signal $s$ to the predator $P$.\\

When we say constant utility we imply that the mutations in the gene vector result in changes in signal but not in utility for the predator. An example would be that the change in the first gene could make a prey more red in color but the first gene does not affect the toxicity or taste of the prey to predator in any way and thus does not influence utility.\\

If utility to predator is proportional to gene vector which in turn is proportional to signal, then utility will increase with increasing signal. Signal can be appropriately scaled along with mutation rate.

\paragraph{Mutation}
Every prey species has a natural system that has evolved with some constraints. There is a cost attributed to most changes. Some changes affect reproduction rate, some affect color, some affect speed or toxicity and other evasion mechanisms. This is a result of mutations in the gene vector. Effectively mutations in the gene vector result in new signals and utilities. \\ 

Mutations can be Gaussian,  which implies that the most of the mutations will be concentrated around the point of origin with a very small probability of large mutations. We consider a simplified version where mutations are restricted to the current signal and the neighbouring two signals. Given mutation rate $p$, then the probability of signal $s$ mutating to $s-1$ or $s+1$ is $p$, while the probability of staying at $s$ is $1 - 2p$.\\

For example, if we consider $u(s) = s$ then it means that a mutation in the gene vector will result in a  change in $s$ and a proportional change in utility.

\paragraph{Signaling distribution}
Every prey has a signal. So a  prey species will have a signaling distribution which is a collection of signals from all the prey of the species. As the species are naturally selected the signal distribution changes shape. this will be described in detail below.

\subsection{Predator}
A predator has the following properties.
\begin{itemize}
    \item Prior Knowledge
    \item Learning system and related parameters
    \item Signal perception function
    \item Utility function
    \item Consumption rate
\end{itemize}
\paragraph{Species}
We also have multiple species of predators each with their own learning systems, utility and perception.

\paragraph{Utility}
The utility function takes in the gene vector of prey as input and return utility to predator for consuming prey as output. As defined in the Prey section, we use the notation $u_{AP}(s)$ which refers to the utility of prey $A$ with signal $s$ to the predator $P$.

\begin{figure}[h]
  \begin{center}
    \subfigure[Actual signal distribution where each color corresponds to a separate prey species.]{\includegraphics[scale=0.45]{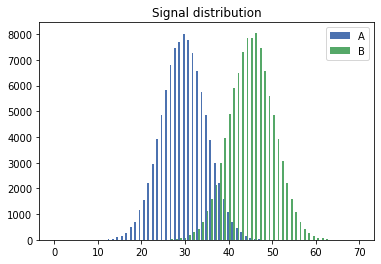}}
    \subfigure[The signal distribution the predator sees as it cannot distinguish between the two prey species.]{\includegraphics[scale=0.45]{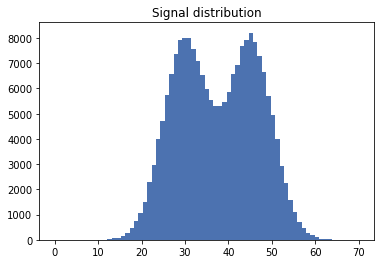}} \\
  \end{center}
  \caption{The two plots contrast difference between actual signal distribution of prey(a) vs the the perceived signal distribution of the predator(b). }
  \label{fig:predator_view}
\end{figure}

\paragraph{Signal from predator perspective}
There are multiple prey species, each with their own signal distribution. The predator only sees the combined distribution of signals. An example has been illustrated in Figure ~\ref{fig:predator_view}.

\paragraph{Signal perception function}
In higher dimensions, predators see a projection of the signal onto a smaller space. For example given a 2 dimensional signal $s=(s_1,s_2)$ of a prey species $A$, a predator $P$ that sees only the first dimension will not be able to distinguish $(10, 3)$ and $(10, 20)$ but it will be able to distinguish between $(10,3)$ and $(5,3)$. To encode this property we use a signal perception function. A real world example is that off color blind predators. Another example is a species that can has a weak smell and weak eyesight and thus distinguishes a prey as a function of both.\\

So a signal perception function $f(s_1,s_2)=s_1+s_2$ will not be able to distinguish between $(10,20)$ and $(25,5)$. This essentially reduces a 2 dimensional signal to a one dimensional signal. This arrangement is sometimes more efficient due to the cost of evolving more complex senses or a more robust learning system.

\paragraph{Consumption Rate} This refers to the number of prey the predator needs to consume at a given day. This can also be easily modified to required amount of utility instead of a required number of prey.

\paragraph{Learning and Prior Knowledge}
We assume that a new predator with no prior knowledge is introduced into a new environment where it has to learn the utilities from the signal.\\

Due to evolution, the signal distributions must be modeled as non-stationary and utility corresponding to a signal keeps changing. Thus the predator needs to forget over time and thus needs a discounting parameter. Clearly neighboring signals will have similar utilities for simple functions. Thus the predator needs to update its learning from a specific signal to surrounding signals as well. This further helps stabilise the learning for sparse signals. Finally we need a predator that explores for new opportunities but also exploits as is naturally seen.

\paragraph{UCB (Upper Confidence Bounded) Bandit}
We consider a standard discounted UCB agent with a bandit for every signal in the range $[L,R]$.[5] During update, the bandits of surrounding signals are also updated. The predator chooses the signal $s$ that maximizes the following value $v_s$.
\begin{center}$ v_s = \mu_s + \sqrt{\left(\frac{\alpha log(n)}{2n_s}\right)}, \quad \quad s \in [L,R]$\end{center}
Where,
\begin{itemize}
\setlength\itemsep{0em}
  \item $\mu_s$ is the discounted average utility gotten so far from signal s and its neighbours.
  \item $\alpha$ is a scaling constant
  \item $n$ is the total number of prey consumed so far
  \item $n_s$ corresponds to the number of prey of signal s consumed so far
\end{itemize}

%As there is no function defined on the trees any more, we need to modify our parameter slightly. 

\section{Signal Drift}
As predators start exploring different signals they learn the corresponding utility over time and start exploiting signals of high utility. This evolutionary game uses essentially natural selection (with replicator dynamics) and results in the drift of the signal distribution for the  particular prey species. But there are various conditions that result in the magnitude and direction of the drift. Remainder of the paper classifies the major ones. \\

There are two main types of signal drift. Signal drift of a prey species due to a signaling range corresponding to a lower utility. The other one being signal drift due to a signaling range with lower expected utility. Depending on the ecosystem one may dominate over the other producing interesting phenomena.

\subsection{Signal Drift to lower prey utility}
\begin{figure}[h]
    \centering
    \includegraphics[trim={0 0 0 0},clip,width=0.8\textwidth]{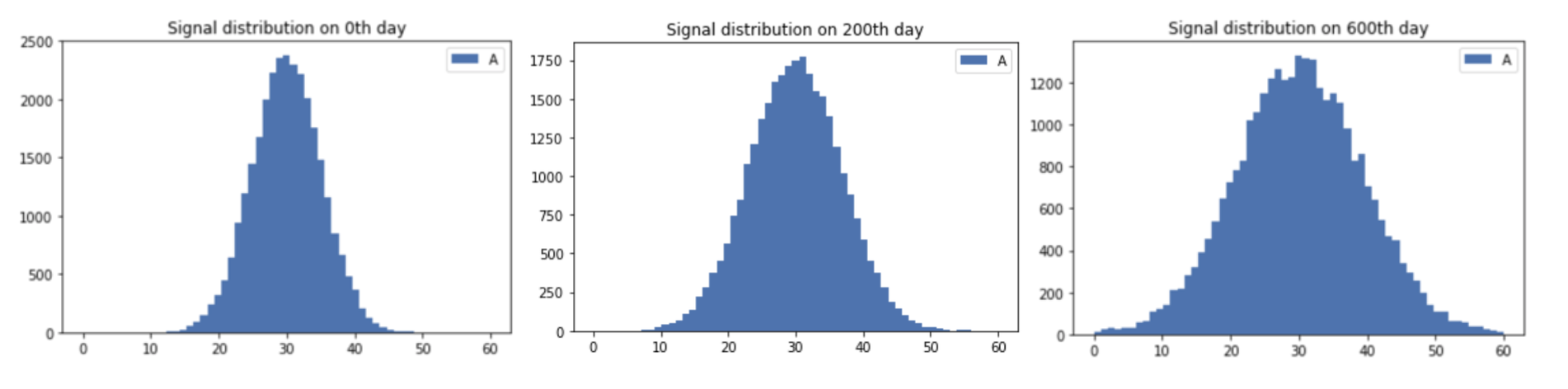}
    \caption{Signal distribution over a 600 generations plotted at different time intervals. The distribution gets flatter and fatter over time. Int this specific example, the rate of consumption by predators is exactly equal to the reproduction rate so the total number of prey at any given generation is approximately 30000.}
    \label{fig:flatten}
\end{figure}

Consider a prey species $A$ with constant utility to predator $P$, meaning that the utility does not change with signal. For example a prey could change its voice frequency but its utility to the predator might not change at all. This situation will result in the predator sampling the prey uniformly, since it prefers every signal exactly the same. An example of which has been shown in Figure ~\ref{fig:flatten}\\
\\

\begin{figure}[h]
    \centering
    \includegraphics[trim={0 0 0 0},clip,width=1\textwidth]{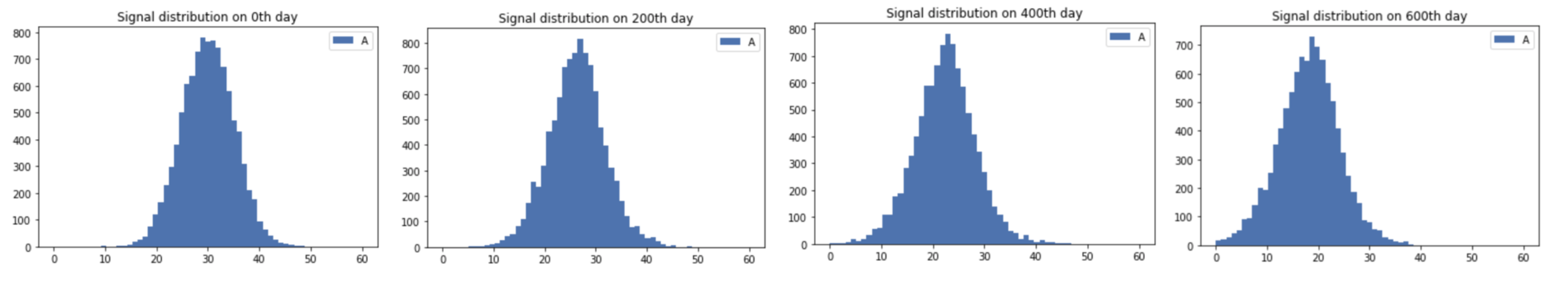}
    \caption{Signal distribution over a 600 generations plotted at different time intervals. We see that the signal distribution shifts left as right has higher utility.}
    \label{fig:right_shift}
\end{figure}

We model utility as a function of signal. Given a signal distribution for the prey species, the prey species mutate and slowly change the signaling distribution. The predator will choose signals that induce a higher utility in the prey. This process results in the movement of the signaling distribution in the other direction. An example of such a adrift is shown in Figure ~\ref{fig:right_shift}. We consider a prey with utility proportional to the signal resulting in a left shift (lower end of the signalling range) because the utility on the left will be lower. \\

\begin{figure}[h]
    \centering
    \includegraphics[trim={0 0 0 0},clip,width=0.8\textwidth]{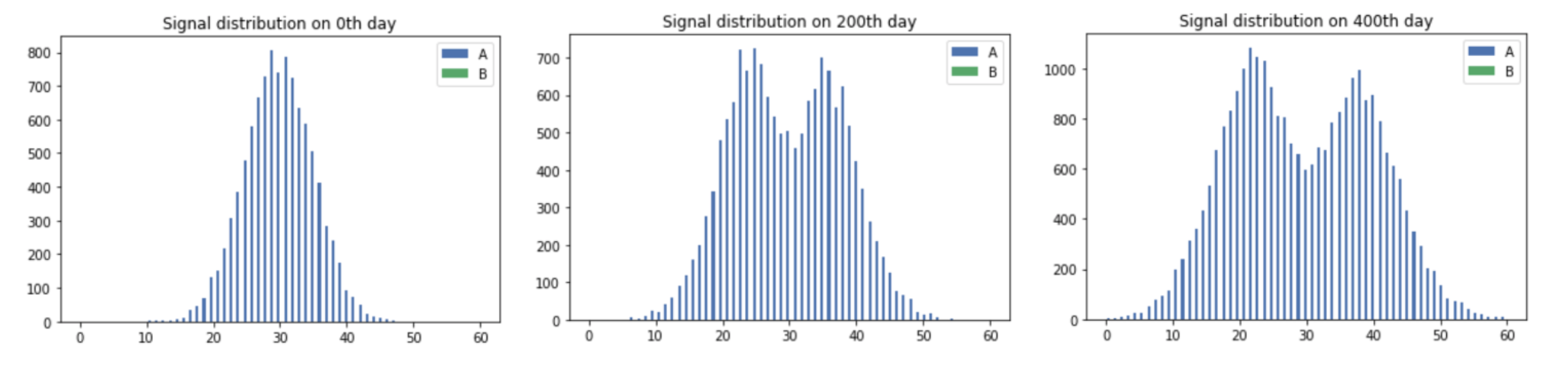}
    \caption{Signal distribution over a 600 generations plotted at different time intervals. The utility function is $u(s) = k - (s-30)^2$ see that the signal distribution splits at the point s=30.}
    \label{fig:non_monotonic}
\end{figure}

An uniform movement in on direction specifically happens in increasing or decreasing functions. If the utility (which is a function of signal) is non monotonic (neither increasing nor decreasing) then the prey population will greedily descend in the direction of lower utility. Figure ~\ref{fig:non_monotonic} illustrates an example with a non monotonic utility function. \\

\subsection{Signal drift due to lower expected utility}
Consider two prey species $A$, $B$ and a predator $P$.\\
$A$(Blue) :Constant high utility to $P$\\
$B$(Green) :Constant low utility to $P$

Over generations, due to exploiting of prey with highest utility by predator we see a natural shift in the signaling range of the population.\\

\begin{figure}[h]
  \begin{center}
    \subfigure[Signal distribution over a 1000 generations plotted at different time intervals.]{\includegraphics[scale=0.35]{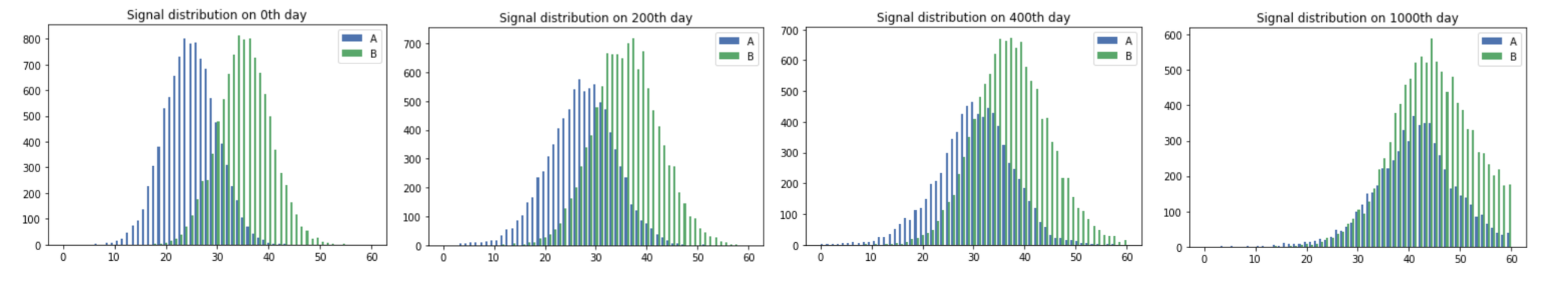}}
    \subfigure[Population time series.]{\includegraphics[scale=0.3]{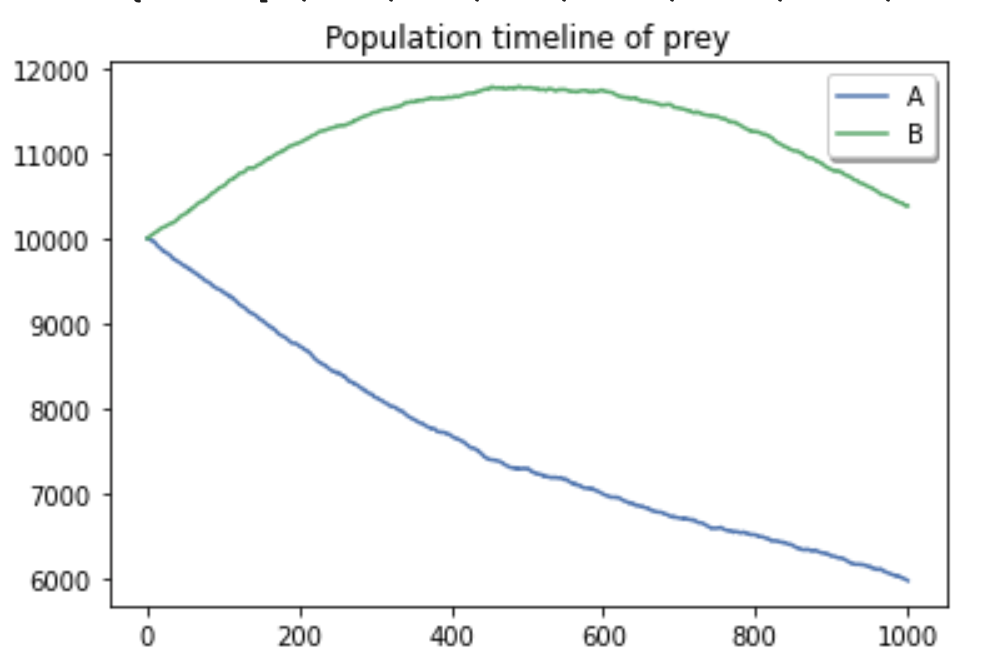}} \\
  \end{center}
  \caption{The plot shows species A(Blue) which is favoured by the predator is predated upon resulting in a right drift of the signal distribution as shown in subplot (a). In this specific example, the total consumption of predator is higher then the growth rate of prey species combined. This results in a  decline of total population over time captured by subplot (b).}
  \label{fig:drift}
\end{figure}

The predator preys on the signaling range with maximum utility. Thus signals that are produced by only the species with high utility will be the prime targets for the predator. On exploitation the prey species with high utility will naturally evolve to signaling range that is shared with a prey species of lower utility. The predator will learn that the common signaling range and associate it with the expected utility. Over generations, a drift in the signal is achieved (Figure ~\ref{fig:drift}). 

Under the same notion, a prey species with low utility will drift away from prey species with higher utility. But this drift will be slower than the drift in case of a prey species with high utility. 

\section{Interesting Phenomena in 1-Dimensional signalling}
We consider a one dimensional gene vector for the prey which corresponds to a one dimensional signal. All species can signal on a band from $L$ to $R$.

\subsection{Pooling Equilibrium}
This phenomena occurs when multiple prey species pool together which results in the predator not being able to distinguish between any of the species. This strategy results in an equilibrium as no species can be naturally selected and thus there is no induced drift.\\

Consider 3 prey species $A$,$B$, $C$ and predator $P$.\\
$A$ (Blue): constant high utility to $P$\\
$B$ (Green): constant low utility to $P$\\
$C$ (Red): constant medium utility to $P$
\begin{center}$u_{BP}(s)<u_{CP}(s)<u_{AP}(s)$\end{center}
\begin{center}$\alpha_B=\alpha_C=\alpha_A$\end{center}

\begin{figure}[h]
    \centering
    \includegraphics[trim={0 0 0 0},clip,width=1\textwidth]{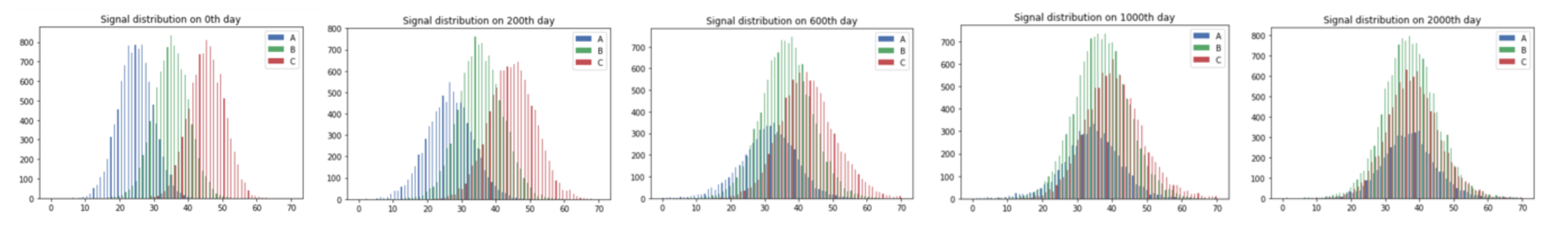}
    \caption{This results in a stable Pooling equilibrium of the three prey species.}
    \label{fig:pooling_eq}
\end{figure}

As seen in Figure ~\ref{fig:pooling_eq} the prey species enter a stable pooling equilibrium. This equilibrium is stable and persists because any change in the signal distribution of a single species will change the distribution of expected utility. This effectively will trigger a drift of all prey species such that the equilibrium is maintained.

\subsection{Multi-peak Mimicry}
This phenomena is the splitting of a single peaked signal distribution of a prey species to a multi-peak distribution. This dynamics can easily be seen in the case of non monotonic functions. Interestingly it is possible to generate the same phenomena using only simple linear monotonic functions.\\

We consider a situation where prey choose a trade-off between reproduction rate and toxicity.
Higher toxicity implies a lower reproduction rate but also a  lower utility to the predator.

Consider 3 prey species $A$, $B$, $C$ and predator $P$.\\
$A$ (Blue): Toxicity is high, utility to predator $P$ is low and reproduction rate is low\\
$B$ (Green):Toxicity is low, utility to predator $P$ is high and reproduction rate is high \\
$C$ (Red): Toxicity is moderate, utility to predator $P$ is moderate and reproduction rate is moderate
\begin{center}$u_{BP}(s)>u_{CP}(s)>u_{AP}(s)$\end{center}
\begin{center}$\alpha_B<\alpha_C<\alpha_A$\end{center}

\begin{figure}[h]
  \begin{center}
    \subfigure[Signal distribution over the generations.]{\includegraphics[scale=0.45]{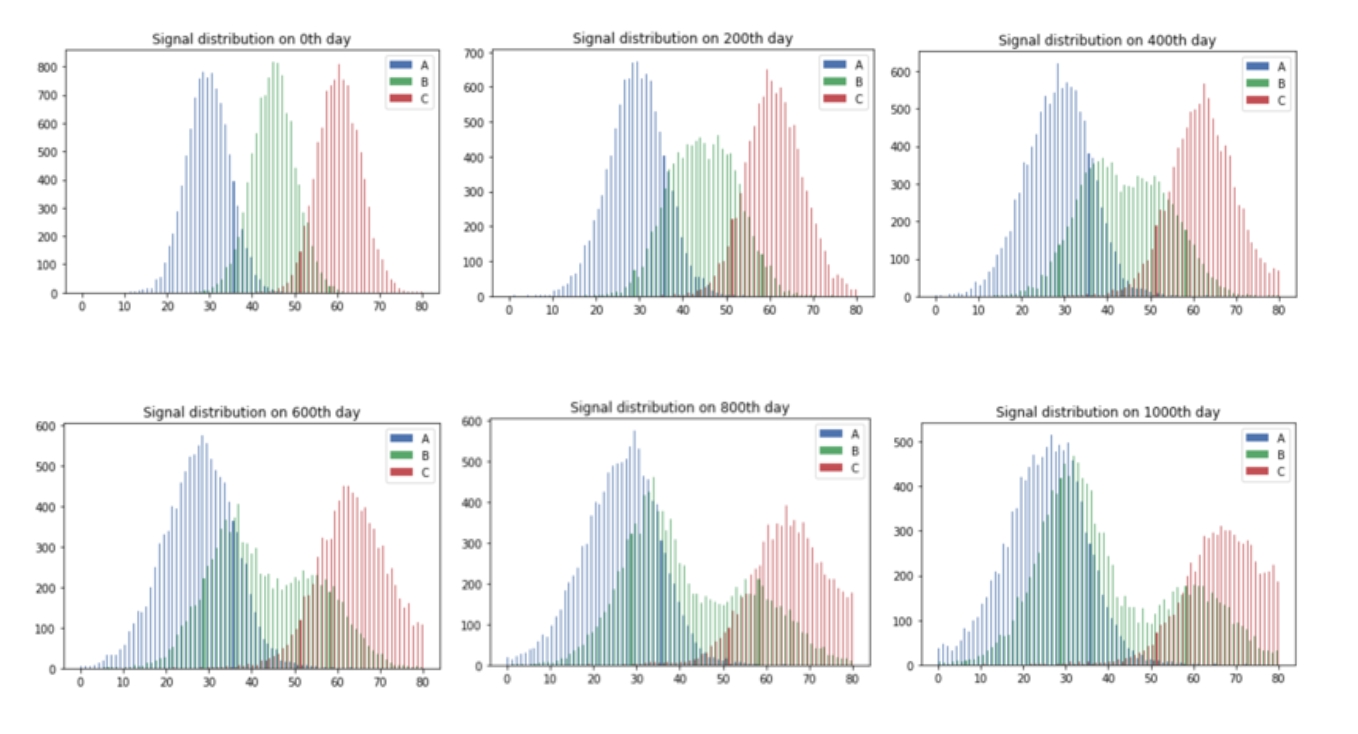}}
    \subfigure[Population time series.]{\includegraphics[scale=0.45]{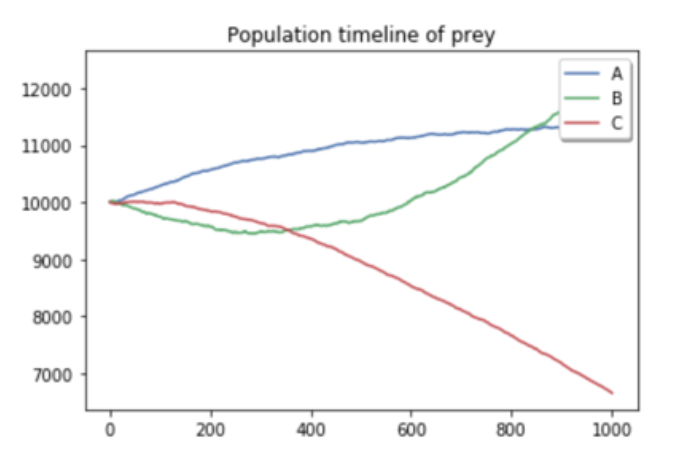}} \\
  \end{center}
  \caption{Initially prey species B is predated upon due to it's utility. This results in the prey species B splitting it's signal and drifting into two separate peaks one mimicking species A and the other peak mimicking species B. Subplot (a) illustrates the changes in the signal distribution while subplot (b) shows the corresponding changes in population of the prey species.}
  \label{fig:multi_peak}
\end{figure}

In the scenario (Figure ~\ref{fig:multi_peak}), prey species $B$ would have gone extinct. But instead we observe that in the presence of mimicry not only does prey species $B$ thrive at the end but also its signal splits into two peaks.

\subsection{Mimicry can be bad to both species}
We show a situation where the mimicry of $A$ and $B$ results in mutual extinction. In the control ecosystem where the other does not exist, they actually thrive.\\

Consider 3 prey species $A$, $B$, $C$ and predator $P$.\\
$A$ (Blue): Utility to predator $P$ is negatively correlated to signal : $(u_{AP}(s) = a_0 - a_1s) $ \\
$B$ (Green): Utility to predator is constant and rate of mutation is insignificant : $(u_{BP}(s) = b_0) $\\
$C$ (Red): Utility to predator is correlated to signal : $(u_{CP}(s) = c_0 + c_1s) $\\

Since utility of $A$ is negatively correlated with signal it drifts towards the right ($R$) because the predator exploits $A$ of lower signal.
Since utility of $C$ is correlated with signal it drifts towards the left ($L$) because the predator selects species $C$ of higher signal. Since, utility of $B$ is constant there is no drift but instead just a flattening of the signal distribution.

\begin{figure}[h]
    \centering
    \includegraphics[trim={0 0 0 0},clip,width=0.7\textwidth]{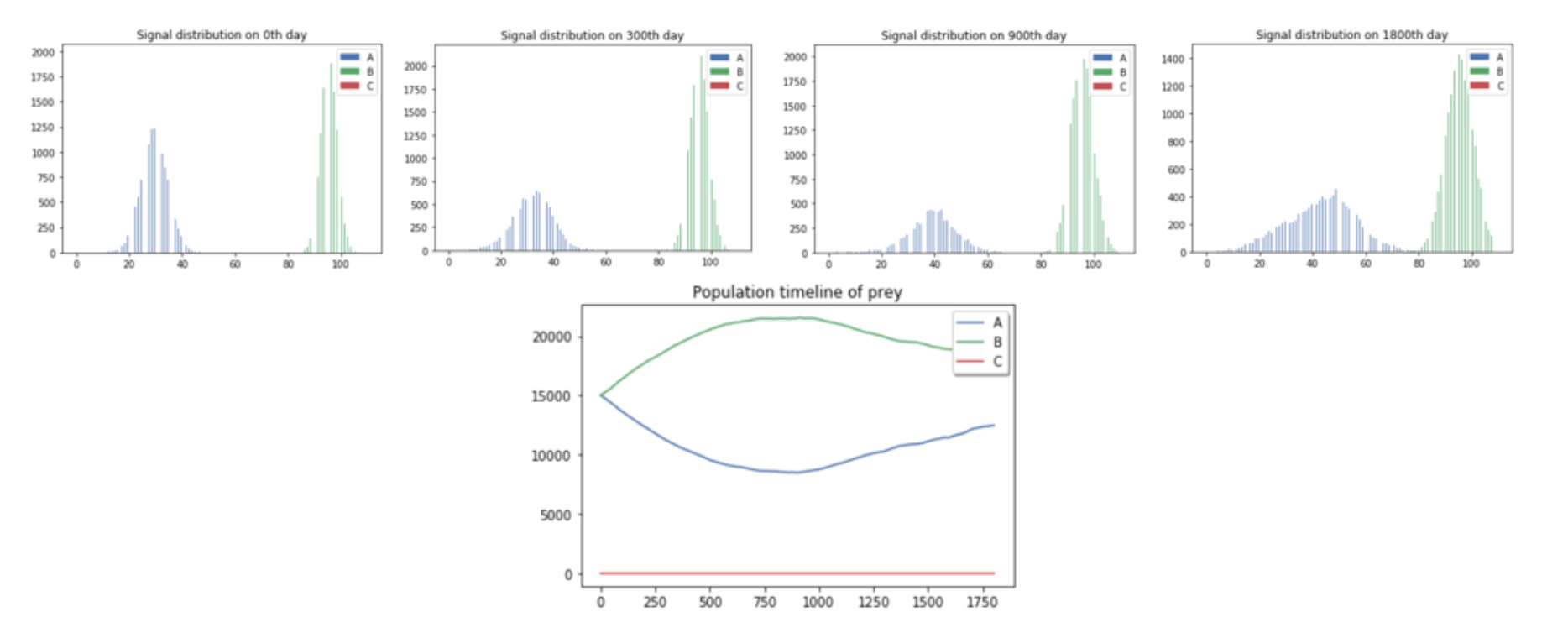}
    \caption{Right drift in signal distribution for species A(Blue) according to the conditions presented in Case 1.}
    \label{fig:bad1}
\end{figure}

\paragraph{Case 1: An ecosystem with only prey species $A$ and $B$:} $A$ is naturally selected towards a signal of lower utility and thus drifts right ($R$). Initially the utility of $A$ is higher and thus the corresponding signals are exploited but as $A$ drifts right its utility decreases till a tipping point where its utility equals the utility of the agents of species $B$. As a result we see an initial drop in population for $A$ followed by recovery which is shown in Figure ~\ref{fig:bad1}.

The turning point is when a critical chunk of species $A$ drifts past to signal higher than $s = \frac{a_0-b_0}{a_1}$. Thus effectively resulting in a lower utility for the predator. A rational predator will make the choice to choose signals corresponding to $B$ compared to $A$.
\\

\begin{figure}[h]
    \centering
    \includegraphics[trim={0 0 0 0},clip,width=0.7\textwidth]{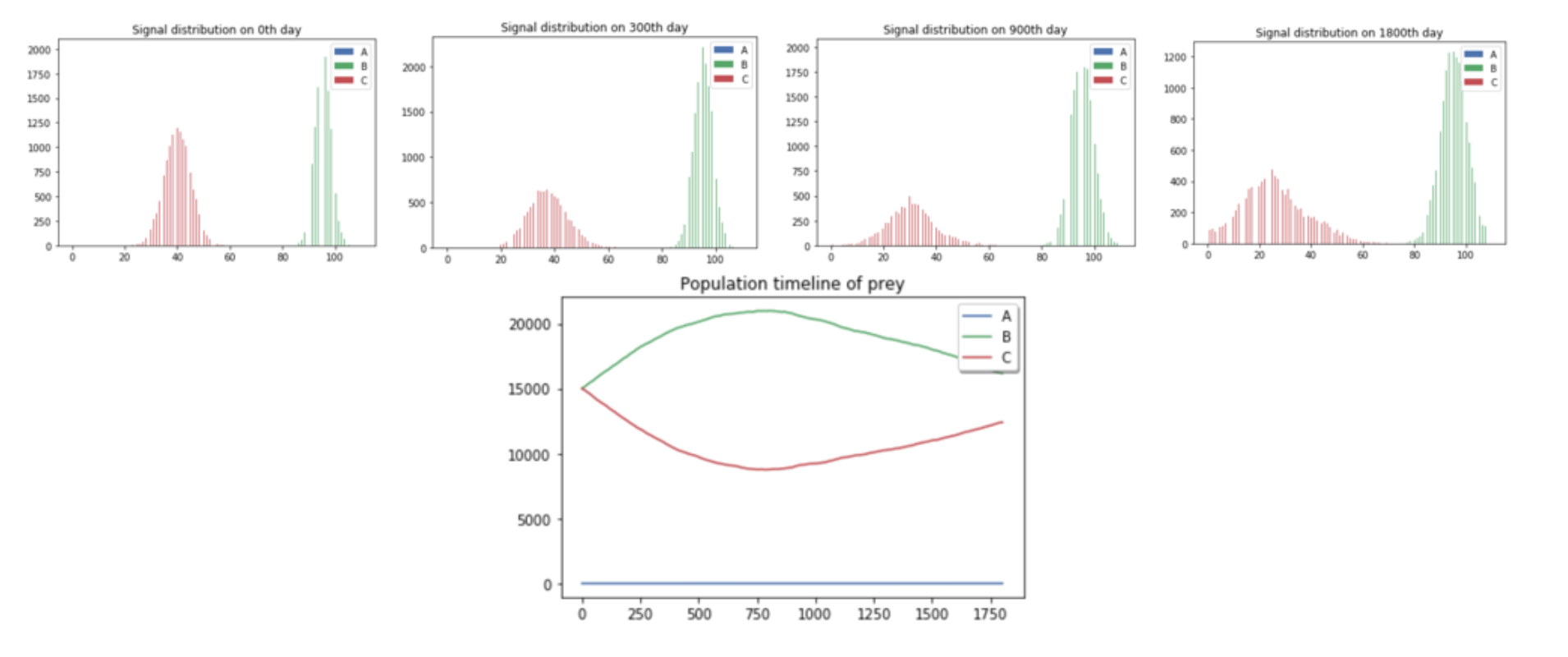}
    \caption{Left drift in signal distribution for species C(Red) according to the conditions presented in Case 2.}
    \label{fig:bad2}
\end{figure}

\paragraph{Case 2: An ecosystem with only prey species $B$ and $C$:} This is very similar to the previous case.
$C$ is naturally selected towards a signal of lower utility and thus drifts left ($L$). Initially the utility of $C$ is higher and thus the corresponding signals are exploited but as $C$ drifts left its utility decreases till a tipping point where its utility equals the utility of the agents of species $B$. As a result we see an initial drop in population for $C$ followed by recovery as shown in ~\ref{fig:bad2}.The turning point is when a critical chunk of species $A$ drifts past to signal lower than $s = \frac{c_0-b_0}{c_1}$.
\\
\\

\paragraph{Case 3: An ecosystem with all prey species $A$,$B$,$C$:}

As a result of the pooling equilibrium formed by the signals of $A$ and $C$. The predator $P$ cannot distinguish between the signals of $A$ and $C$ and thus cannot naturally select them.\\

At the point of pooling, if the expected utility of the signals in the pooling equilibrium is lesser than the utility of the signals of prey species $B$. This leads to exploitation and thus eventually extinction.\\

\begin{figure}[h]
    \centering
    \includegraphics[trim={0 0 0 0},clip,width=0.5\textwidth]{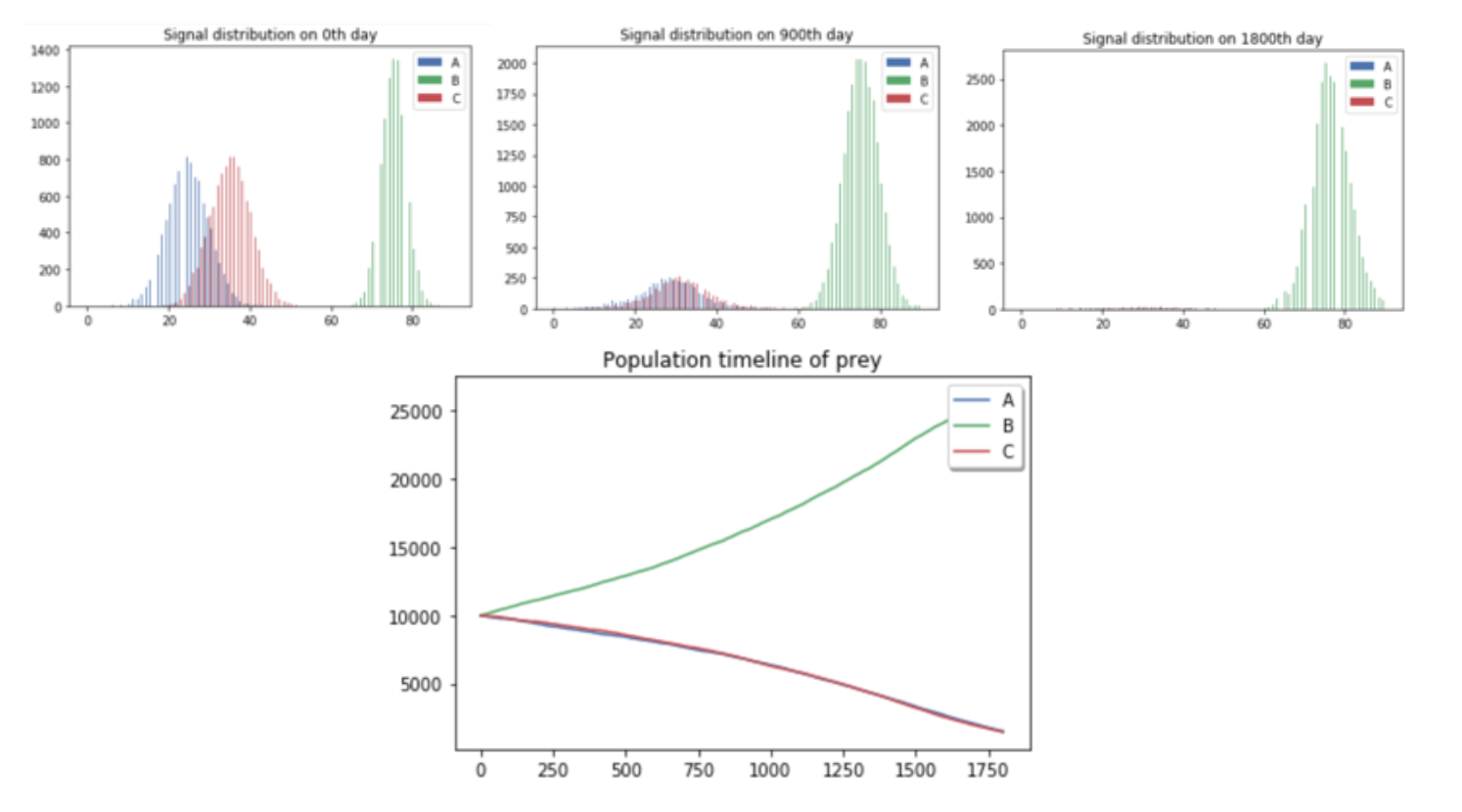}
    \caption{In a combined setting of all 3 prey species A,B and C. We see that the drift of species A and C result in a pooling equilibrium resulting in the extinction of both species. We considered $u_{AP}(s) = 3 - 0.05s, u_{BP}(s) =1.3, u_C(S)=0.05s$ and initial distribution means at points 25 for A and 35 for C. $s=30$ will be the mean of the pooling equilibrium between $A$ and $C$ after drifting. This dynamics is because initially both $A$ and $C$ have the same utility and are thus selected together  The expected utility of any signal in this pooling equilibrium will be 3/2. This is greater than the utility of prey $B$ and thus the pooling equilibrium will be exploited till extinction.}
    \label{fig:bad3}
\end{figure}

Figure ~\ref{fig:bad3} represents Case 3 which is a combination of both Case 1 and 2. Although a specific example, it can be easily generalized to show that a pooling equilibrium can form. Thus resulting in the stopping of natural selection towards a lower utility, effectively leading to extinction. This extinction is mainly because of the lack of freedom in the 1-Dimensional signaling space.

\subsection{Oscillating Mimicry}
Oscillating mimicry where signals of two species is constantly oscillating, is not possible when we consider a one dimensional gene vector, identity signal function ($g$) and a fixed predator population. This situation is due to the formation of a pooling equilibrium.\\

\section{Interesting Phenomena in 2-Dimensional signalling}
2-Dimensional signaling opens up a whole new world phenomena. In 1-Dimensional signaling due to a lack of freedom we are restricted to certain type of phenomena. here we present two basic ideas of how far we can go when we increase the signalling dimension.

\subsection{Drift in 2-Dimensions}
Drift in higher dimensions is similar to 1-dimensional case except with more freedom.The drift is determined by the signal perception function, utility function of predator and genetic vector of prey.\\

\begin{figure}[h]
    \centering
    \includegraphics[trim={0 0 0 0},clip,width=0.7\textwidth]{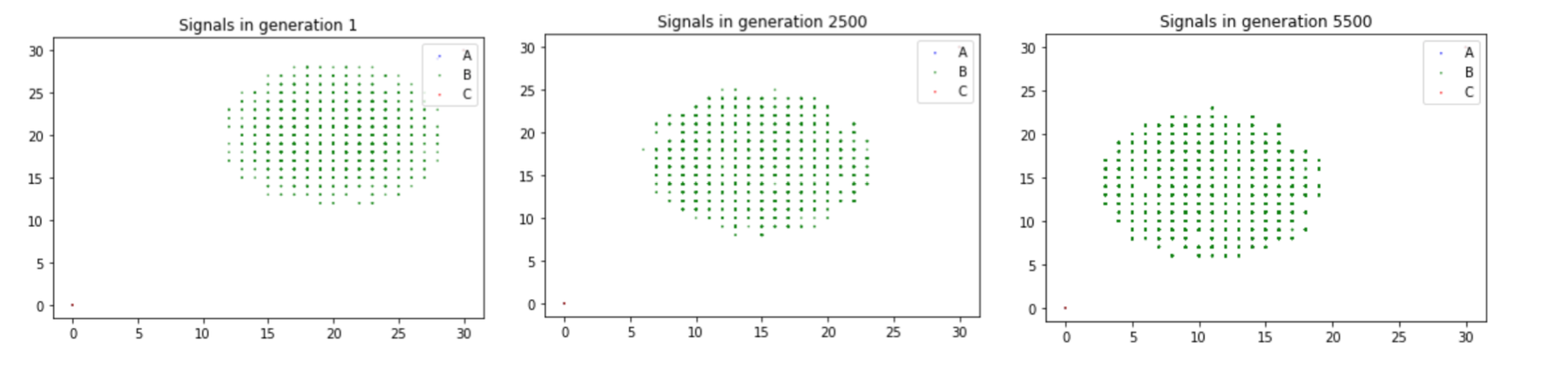}
    \caption{Drift in the 2-dimensional signal distribution of a prey species perpendicular to the predator's signal perception function.}
    \label{fig:2D}
\end{figure}

Consider a prey species $A$ and a predator $P$. $A$ has a genetic constraint that it cannot reproduce if its genes are too different from the rest. This constraint stops the the signaling distribution from spreading out in all directions. The predator $P$ can see signals only along the line $s_1 - s_2 = k$ because of the signal perception function $f(s_1,s_2) = s_1+s_2$. Furthermore a utility function $u_{AP}(s_1,s_2) =s_1+s_2$ results in a diagonally left-down drift for the prey species $A$. The resulting drift of a 2-dimensional signal distribution has been illustrated in Figure ~\ref{fig:2D}

\subsection{Switching directions}
In the presence of two prey species and two predator species, we can create very interesting phenomena which involve changes in drift direction. Such directional change is not possible in 1-Dimensional signaling due to the creation of a Pooling Equilibrium.\\

Consider two prey species $A$, $B$ and two predator species $P$ and $Q$. Let the predators and prey have the following functions.
\begin{itemize}
    \item $P$ signal perception function : $f_P(s_1,s_2)=s_1$. This subcase means that $P$ can distinguish only the $x$ coordinate of the signal.
    \item $Q$ signal perception function : $f_Q(s_1,s_2)=s_2$. This subcase means that $Q$ can only distinguish the $y$ coordinate of the signal.
    \item The signaling function of $A$ as a function of the gene vector : $g_A(v_1,v_2)=(v_2,-v_1)$
    \item The signaling function of $B$ as a function of the gene vector : $g_B(v_1,v_2)=(-v_1,-v_2)$
    \item The utility function of $P$ as a function of gene vector : $u(v_1,v_2) \propto v_1+v_2$
    \item The utility function of $Q$ as a function of gene vector : $u(v_1,v_2) \propto v_1-v_2$
\end{itemize}
The above functions result in the following utility functions.
\begin{itemize}
    \item $u_{AP}(s_1,s_2)\propto s_1-s_2$
    \item $u_{AQ}(s_1,s_2)\propto -s_1-s_2$
    \item $u_{BP}(s_1,s_2)\propto -s_1-s_2$
    \item $u_{BQ}(s_1,s_2)\propto -s_1+s_2$
\end{itemize}
Starting at the appropriate positions with correct scaling parameters such that species $A$ has all the $x$ coordinates of its signal less then $B$ but $y$ coordinates of its signal greater than $B$. If initially $u_{AP} > u_{BP}$ and $u_{BQ} > u_{AQ}$, then $A$ will drift left but to predation by $P$ and $B$ will drift down due to predation by $Q$. The drift results in a mean x coordinate decrease for species A and a mean y coordinate decrease in signal for species B. As shown by the utility function the drift in species A results in decrease of $u_{AP}$ but an increase in $u_{AQ}$. Similarly, the drift in species B results in decrease of $u_{BQ}$ but an increase in $u_{BP}$. Thus after a point when $u_{AP} = u_{BP}$ and $u_{BQ} = u_{AQ}$, the predators start gaining interest in the other prey which they were not predating originally. Slowly the predators reduce the predating of their original prey and begin predating the other prey. This will result in $A$ drifting upward due to predation by $Q$ and $B$ drifting right due to predation by $P$.\\

So $A$ initially starts drifting left, followed by a gradual change in direction resulting in an upward drift. Similarly $B$ initially drifts downward and gradually shifts to drifting right.

\section{Concluding Remarks}
\label{sec:conclusion}

In most of the phenomena we have focused on using only linear functions. Although this may not be the case, it only shows that a simple set of linear rules can result in complex mechanisms like John Conway's ``The Game of Life.'' If we introduce higher degree or more complex functions we will be able to construct even more fascinating phenomena for bio-mimicry. Furthermore if we increase the dimension of signaling and the nature of genetic constraints, we can explore an even wider variety of counter-intuitive results.\\

This paper gives us a broad framework to understanding the existence of multiple mathematically feasible mimicry based phenomena. This can easily be extended to the dynamics between multiple agents that can systematically be described with learning predators and evolving prey.\\

This framework can be used is in understanding the introduction of Artificial intelligence or learning systems into new ecosystems. %Another example is that of fake currency which is a mimic of real currency. The predators will be the verifiers who learn over time to detect fake currency better while malicious agents work on new and improved methods to evade the verifiers. If we design a set of learning bots that learn to remove fake currency from a system. The bots learn over time and utility is precisely that of removing fake currency and not real currency. Fake currency can be modelled as prey. If fake currency is left unchecked it will be misused and there will be more malicious agents pumping it into the economy without the fear of getting caught. Furthermore too much fake currency also does not work and this represents prey-prey competition. If we identify enough fake currency we can narrow down the source which is similar to lowering reproduction in prey due to a smaller population.\\%
An example is the case of malicious bots getting better over time while security systems learn to deal with them. Apart from this, multiple other systems can effectively be modeled using this framework. By understanding these dynamics better we can potentially stop certain types of behaviour if required, by introducing minimal changes.

\section{References}

\begin{enumerate}
    \item Rowland HM, Fulford AJT, Ruxton GD. 2017. Predator learning differences affect the survival of chemically defended prey. Animal Behaviour. 124:65–74.\\ doi:10.1016/j.anbehav.2016.11.029. \\
    \item Skelhorn J, Halpin CG, Rowe C. 2016. Learning about aposematic prey. Behavioral Ecology. 27(4):955–964. doi:10.1093/beheco/arw009. \\
    \item Müller, Fritz (1879). "Ituna and Thyridia; a remarkable case of mimicry in butterflies. (R. Meldola translation)". Proclamations of the Entomological Society of London. 1879: 20–29.\\
    \item Enaganti I, Mishra B. 2021 Mar 25. Lotka-Volterra Equations in the Presence of Mimicry. doi:10.1101/2021.03.25.436931. \\
    \item Garivier A, Moulines E. 2008 May 22. On Upper-Confidence Bound Policies for Non-Stationary Bandit Problems. arXiv:08053415 [math, stat]. [accessed 2021 Apr 17].\\ https://arxiv.org/abs/0805.341\\
\end{enumerate}

\end{document}